\documentclass[twocolumn,showpacs,preprintnumbers,amsmath,amssymb,aps,superscriptaddress]{revtex4}
\usepackage{graphicx}
\begin{document}
%\title{Quantum gate using qubit states separated by a terahertz}
\title{Quantum gate using qubit states separated by terahertz}

\author{K. Toyoda}
\affiliation{
Graduate School of Engineering Science, Osaka University, 
1-3 Machikaneyama, Toyonaka, Osaka 560-8531, Japan
}
\affiliation{
JST-CREST, 4-1-8 Honmachi, Kawaguchi, Saitama 331-0012, Japan
}
\author{S. Haze}
\affiliation{
Graduate School of Engineering Science, Osaka University, 
1-3 Machikaneyama, Toyonaka, Osaka 560-8531, Japan 
}
\author{R. Yamazaki}
\affiliation{
JST-CREST, 4-1-8 Honmachi, Kawaguchi, Saitama 331-0012, Japan
}
\author{S. Urabe}
\affiliation{
Graduate School of Engineering Science, Osaka University, 
1-3 Machikaneyama, Toyonaka, Osaka 560-8531, Japan
}
\affiliation{
JST-CREST, 4-1-8 Honmachi, Kawaguchi, Saitama 331-0012, Japan
}

%             %  but any date may be explicitly specified

\begin{abstract}
A two-qubit quantum gate is 
realized using electronic excited states in a single ion 
with an energy separation on the order of
a terahertz times the Planck constant as a qubit.
Two phase locked lasers are used to excite a stimulated Raman transition
between two metastable states $D_{3/2}$ and $D_{5/2}$ separated by 1.82 THz
in a single trapped $^{40}$Ca$^+$ ion to construct a qubit,
which is used as the target bit
for the Cirac-Zoller two-qubit controlled NOT gate.
Quantum dynamics conditioned on a motional qubit 
is clearly observed as a fringe reversal in Ramsey interferometry.
\end{abstract}

\maketitle

%\section{Introduction}

Atomic systems including trapped ions and neutral atoms 
are considered attractive for quantum information processing (QIP) since
they can be made to be well isolated from the environment and 
hence enable construction of qubits with small decoherence/dephasing. 
Among experimental approaches toward QIP using different physical systems,
some of the most advanced have been experiments using trapped ions%
\cite{Blatt2008,Haffner2008},
which are based on 
qubit levels with separation in the rf/microwave region and the optical region.
Recent advances in optical comb generation and optical frequency metrology%
\cite{Diddams2000,Udem2002,Kourogi1993}
offers much flexibility in choosing qubit states,
including atomic states with frequency separations
that have not been explored before.

In view of recent progress in experiments of ultracold molecules
transferred to 
the ground state of both internal and external degrees of freedom,
molecular systems which have rich internal structures
are also considered attractive for application to QIP.
In the recent works\cite{Ni2008,Danzl2008,Lang2008,Deiglmayr2008}, 
by performing stimulated Raman adiabatic passage using two lasers with
high relative coherence,
weakly bound ultracold Feshbach molecules are transferred to
their rovibronic ground state.
In addition, 
there are proposals to encode qubits in molecular states with
small dipole moments and transfer these to states with larger
dipole moments, thereby realizing switchable interaction between
molecular qubits\cite{Carr2009,Lee2005,Yelin2006,Charron2007}. 
The required transfer can be performed by applying
two phase locked lasers through stimulated Raman process.

In this article, we present the result of a quantum gate experiment
using phase-locked lasers to excite a stimulated Raman transition.
Two metastable states $D_{3/2}$ and $D_{5/2}$
in $^{40}$Ca$^+$ separated by 1.82 THz are used as the target bit 
to perform the Cirac-Zoller gate\cite{Cirac1995}.
This is the first attempt to use phase locked lasers 
to bridge an energy separation larger than a terahertz and 
realize a quantum gate,
and is an important step toward
obtaining a wider choice of qubit levels including
internal levels of atoms and molecules.

%\section{Principles}
Cirac and Zoller proposed in 1995 
a realistic scheme for scalable quantum computation 
using a string of ions in a linear trap\cite{Cirac1995}.
It was experimentally demonstrated in a simplified form
using internal states and a motional degree 
of freedom in a single $^9$Be ion\cite{Monroe1995}.
A full implementation of the scheme in a scalable manner 
using a $^{40}$Ca$^+$ ion string with the technique of individual
addressing is reported in 2003\cite{Schmidt-Kaler2003}. 

It has been shown that 
all unitary operations on arbitrary many qubits can be decomposed into
two-bit gates and one-bit gates\cite{Divincenzo1995}.
One example of such decomposition of unitary operations 
uses controlled NOT (CNOT) gates and rotation operations on single qubits%
\cite{Barenco1995a,Barenco1995b}.
Analogously to a classical exclusive-OR (XOR) gate,
a CNOT quantum gate realizes the following operation:
$|\epsilon_1\rangle|\epsilon_2\rangle\rightarrow|\epsilon_1\rangle|\epsilon_1\oplus\epsilon_2\rangle$
with $\epsilon_{1,2}=1,2$ and $\oplus$ representing 
an addition modulo 2.

\begin{figure}[b]
%\begin{figure}[b]
\includegraphics[height=2in]{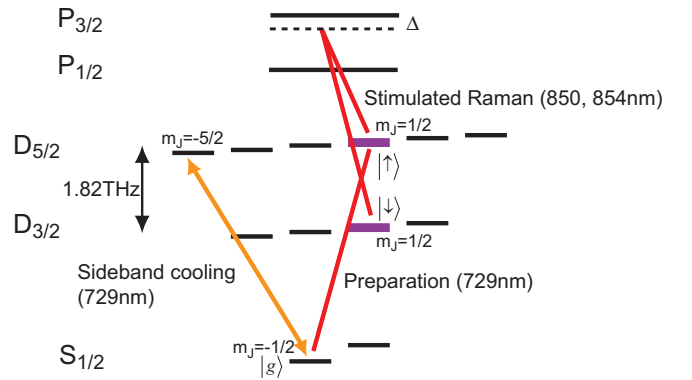}
\caption{%
(Color online)
Level scheme for $^{40}$Ca$^+$ and transitions 
relevant for implementing the CZ gate.
Two sublevels each from $D_{3/2}$ and $D_{5/2}$ metastable states,
$\left|\uparrow\right\rangle\equiv\left|D_{5/2}(m_J=1/2)\right\rangle$ and
$\left|\downarrow\right\rangle\equiv\left|D_{3/2}(m_J=1/2)\right\rangle$,
are used as the qubit states here.
}
\label{fig:levels}
\end{figure}

In the Cirac-Zoller (CZ) proposal\cite{Cirac1995}, 
for implementing this CNOT operation,
a red-sideband pulse, which 
is detuned to the lower side of the resonance of the qubit
transition
by the frequency of a collective motional mode,
is applied between one basis state of the target qubit
and an auxiliary state.
When the collective motional state has one quantum,
the red-sideband pulse applied for a duration corresponding 
to a 2$\pi$ rotation causes 
a $\pi$ phase shift between two basis states of the target
qubit states.
On the other hand, when the collective motional state has no quantum,
such rotation does not occur and no phase shift is given to the target qubit.
This corresponds to a unitary operation 
conditioned on the motional quantum number, 
thereby implementing a controlled phase gate,
and a CNOT gate is realized when this is
combined with certain single qubit operations.

%\section{Experimental procedure}
To realize the CZ gate using the 
metastable states in $^{40}$Ca$^+$ and its motional states,
we adopt an excitation scheme using
the stimulated Raman transition between $D_{3/2}$ and $D_{5/2}$ 
along with a quadrupole
transition that connects the ground state $S_{1/2}$ with $D_{5/2}$
%Figure \ref{fig:levels} shows the relevant levels in $^{40}$Ca$^+$.
(see Fig.\ \ref{fig:levels}).
The stimulated Raman transition is used for single qubit operation
on the metastable states qubit, while the quadrupole transition
is used for realizing conditional phase shift required for a CZ gate,
as well as sideband cooling and state preparation.
As the target qubit states
% for the CZ gate
%one Zeeman component in the $D_{5/2}$ and another in $D_{3/2}$ state
%are chosen as the qubit states, which are denoted as
$\left|\uparrow\right\rangle\equiv{}D_{5/2}(m_J=1/2)$ and
$\left|\downarrow\right\rangle\equiv{}D_{3/2}(m_J=1/2)$
are chosen,
while as the control qubit
the low-lying two states of the axial motion initialized to the
ground state are used:
$\left|0\right\rangle(\left|1\right\rangle)\equiv%
\left|n_z=0\right\rangle(\left|n_z=1\right\rangle)$,
where $n_z$ denotes the axial motional quantum number.
A conditional phase shift is implemented by 
applying a blue sideband (BSB) $2\pi$ pulse 
between $\left|\uparrow\right\rangle$ and
$\left|g\right\rangle\equiv{}S_{1/2}(m_J=-1/2)$ state, 
which gives a $\pi$ phase shift to the 
$\left|1\right\rangle\left|\uparrow\right\rangle$.

%\begin{figure}[t]
\begin{figure}[t]
\includegraphics[height=1.5in]{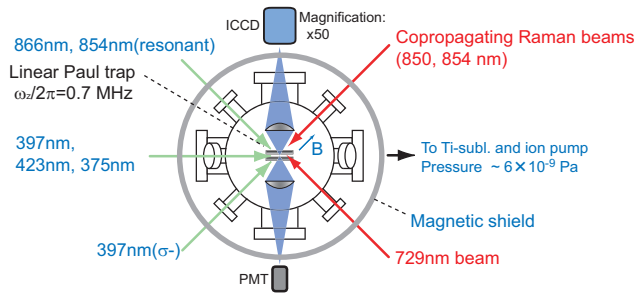}
\caption{%
(Color online)
Experimental setup for the terahertz-qubit quantum-gate experiment.
See text for details.
}
\label{fig:setup}
\end{figure}

%\section{Experimental Setup}
We use a single $^{40}$Ca$^+$ ion 
trapped in a vacuum pressure of $6\times10^{-9}$ Pa.
The trap used here is a conventional linear trap with
an operating frequency of 23 MHz and 
secular frequencies of $(\omega_x,\omega_y,\omega_z)/2\pi=(1.91,1.68,0.72)$
MHz.
A magnetic field of $\sim3.1\times{}10^{-4}$ T is applied to lift the
degeneracy of Zeeman states and to define the quantization axis for 
optical pumping.
To reduce the effects of the ambient ac magnetic field,
the vacuum chamber is enclosed in a magnetic shield.
Loading of single ions is performed by using two-step photoionization 
from the $4s\,^1S_0$ ground state of Ca via $4p\,^1P1$ 
with corresponding wavelengths of 423 nm and 375 nm
for the first and second step of the photoionization, respectively.

About the phase locked lasers used for 
excitation of the stimulated Raman transition,
the setup has been modified from the one described 
in our previous article\cite{Yamazaki2007}
in order to improve the noise in the difference 
frequency of the two lasers.
Two Ti-sapphire lasers at 850 and 854 nm phase locked by using a
passive-type optical comb\cite{Kourogi1993}
in combination with an acousto-optic modulator (AOM) and an
electro-optic modulator\cite{Hall1984}
are used to excite the stimulated Raman transition.
For excitation of the quadrupole transition, 
a titanium sapphire laser at 729 nm
stabilized to a high-finesse low-thermal-expansion cavity
having a linewidth of
$<$ 400 Hz and a root-mean-square intensity noise of 0.3\% is used.
Control of optical frequency/phase/amplitude is done by 
AOM and rf fields used for them are generated by 
direct-digital synthesis (DDS) boards which are controlled by a field-programmable
gate array (FPGA).

See Fig.\ \ref{fig:setup} for the details of the beam configuration.
For realizing gate experiments, all the motional degrees of the ion are
cooled to near the ground states
using Doppler cooling (with 397 and 866 nm lasers) and sideband cooling (SBC).
For SBC, the
$S_{1/2}(m_J=-1/2)$--$D_{5/2}(m_{J'}=-5/2)$ transition at 729 nm 
is used, and an additional quenching laser resonant to
$D_{5/2}$--$P_{3/2}$ at 854 nm is also applied. 
All the three dimensions are cooled for 2 ms each and then this is
repeated for 20 times.
Optical pumping is performed using 397 nm $\sigma^-$ transition in
$S_{1/2}(m_J=+1/2)$--$P_{1/2}(m_{J'}=-1/2)$
before, every 6ms during, and after SBC,
each with duration of 6 $\mu$s.
The final quantum numbers obtained after SBC are
$(\overline{n_x},\overline{n_y},\overline{n_z})\sim(1,1,0.02)$.

For preparation to the $\left|\uparrow\right\rangle$ state,
a carrier/BSB $\pi$ pulse on 
$\left|g\right\rangle$--$\left|\uparrow\right\rangle$ 
is applied. 
This is a $\left|\Delta{}m_J\right|=1$ transition
which requires a polarization different from
that used for sideband cooling transition
for which $\left|\Delta{}m_J\right|=2$.
In our case the former is parallel with and the latter is perpendicular
to the surface of the optical table on which the trap chamber is placed
(see Fig.\ 2).
In order to perform both in one configuration,
the polarization of the 729nm light is chosen to be linear and
rotated from the perpendicular direction
by 45 degree.

The target qubit states are discriminated by shining the cooling lasers
at 397 and 866 nm for 7 ms and observing fluorescence photons by
a photomultiplier.

The coherence times of the Raman transition have been measured
to be $\sim$5.1 ms (1.6 ms) with (without) spin
echo in a setup without a magnetic shield\cite{Toyoda2009}.
The coherence time for the quadrupole transition
with a magnetic shield, which is deduced from
decay of Rabi oscillation signals, is about 0.8 ms.

\begin{figure}[t]
\includegraphics[height=2.5in]{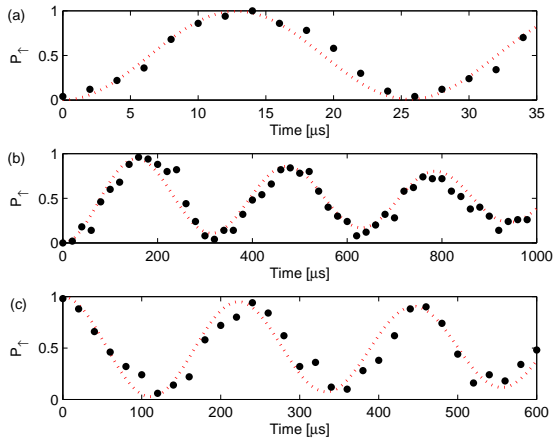}
\caption{%
(Color online)
Rabi oscillation signals
in the transitions 
relevant to the terahertz qubit scheme.
(a) Carrier (BSB) Rabi oscillation and
(b) BSB Rabi oscillation in 
$\left|g\right\rangle$--$\left|\uparrow\right\rangle$.
(c) Carrier Rabi oscillation in the stimulated Raman transition 
$\left|\uparrow\right\rangle$--$\left|\downarrow\right\rangle$.
The dotted curves are results of numerical simulation.
}
\label{fig:rabi}
\end{figure}

%\section{Experimental results}
%\subsection{Rabi oscillation on the relevant transitions}
Figure \ref{fig:rabi} %
shows Rabi oscillation signals on the relevant transitions
including the carrier/BSB on 
$\left|g\right\rangle$--$\left|\uparrow\right\rangle$
and the carrier on the stimulated Raman transition.
Based on these results, 
we can expect nearly unit fidelity for carrier pulses on 
$\left|g\right\rangle$--$\left|\uparrow\right\rangle$
while less fidelity for BSB Rabi pulses and carrier pulses 
on the stimulated Raman transition.
%The factors that limit fidelity are considered later
%in \S\ref{sec:gate_fidelity}.
The figure also shows results of numerical simulation,
%We can extract 
the details of which are given later.
By comparing the simulation with the experiment,
we can quantitatively characterize the fidelity limiting
factors, 
and this helps estimation of possible fidelity of Bell state generation
as described later. % in \S\ref{sec:num_simu},
%where the details and results of simulation is given.

% \subsection{Rabi oscillation on the stimulated Raman transition}

\begin{figure}[b]
\includegraphics[height=3in]{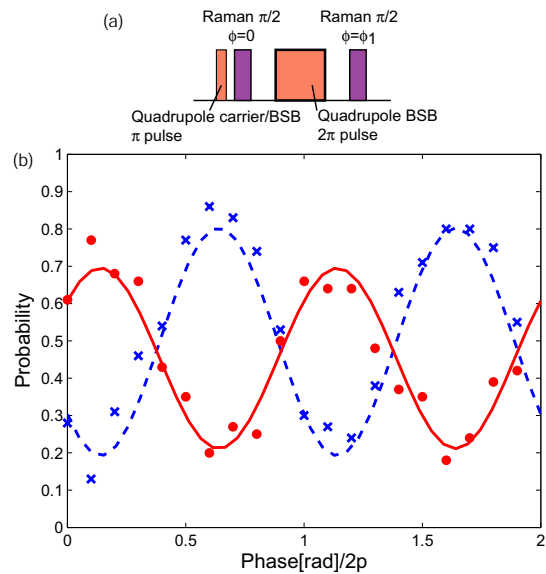}
\caption{%
(Color online)
(a) Pulse sequence for the Cirac-Zoller gate experiment. %
(b) Result of a Cirac-Zoller gate experiment.
Ramsey interference signals are plotted against the phase
of the second pulse.
Crosses (filled circles) represent 
the $\left|0\right\rangle$ ($\left|1\right\rangle$)
preparation case.
The solid (dashed) curve represents numerical
simulation for $\left|0\right\rangle$ ($\left|1\right\rangle$) preparation.
The $\pi$ phase difference between interference fringes indicates that
a controlled dynamics characterizing the Cirac-Zoller gate 
is realized.}
\label{fig:czgate}
\end{figure}

%\subsection{Results for the Cirac-Zoller gate experiment}
Figure \ref{fig:czgate}(a) shows the pulse sequence for the
Cirac-Zoller gate experiment.  
The first pulse ({\it preparation pulse}) 
is applied either on the carrier or BSB on
$\left|g\right\rangle$--$\left|\uparrow\right\rangle$
to prepare the motional state $\left|0\right\rangle$ or $\left|1\right\rangle$ 
respectively.  Then the first stimulated-Raman $\pi/2$ pulse 
is applied, which is followed by a BSB $2\pi$ pulse on 
$\left|g\right\rangle$--$\left|\uparrow\right\rangle$
and the second stimulated-Raman $\pi/2$ pulse.
For $\left|0\right\rangle$ preparation the BSB $2\pi$ pulse
cause no effect since there is no motional state to reach in 
$\left|g\right\rangle$, while 
for $\left|1\right\rangle$ preparation
the BSB $2\pi$ pulse cause
$2\pi$ rotation and gives a $\pi$ phase shift to the original state.
This conditional phase flip 
($\pi$ rotation around the $z$ axis in the Bloch sphere)
is converted into a conditional bit flip
($\pi$ rotation around a horizontal axis in the Bloch sphere)
by the two $\pi/2$ pulses.

%\begin{figure}[t]

Figure \ref{fig:czgate}(b) shows the result of the Cirac-Zoller gate
experiment.
Here the phase of the second pulse is rotated from 0 to $4\pi$ and
the population in $\left|\uparrow\right\rangle$ is measured.
Crosses represent the $\left|0\right\rangle$ preparation case, and
filled circles the $\left|1\right\rangle$ preparation case.
The interference fringes for the two cases clearly show a $\pi$
phase difference to each other, which is an evidence of 
a conditional dynamics caused by the BSB $2\pi$ pulse.
The contrasts of the fringes are limited to $0.4\sim0.6$, and
for the BSB preparation case there is 
a negative offset $\sim0.05$, which is consistently observed in 
similar measurements. These imperfections are explained later.

%\begin{figure}[t]

%\subsection{Numerical simulation}
\label{sec:num_simu}
Numerical simulation is performed to quantitatively
analyze the CZ gate result and to estimate possible fidelity for
Bell state generation.
A Liouville equation with exponential decay is solved for three internal levels 
($\left|g\right\rangle$, $\left|\uparrow\right\rangle$ and 
$\left|\downarrow\right\rangle$) and 5 external levels representing
the axial motional states truncated at $n_z=4$.
Coupling between the internal and external states is considered
to the second order of the Lamb-Dicke factor $\eta$ 
for carrier excitation and to the first order for sideband excitation. 

Fidelity limiting factors except for that 
from axial motional state distribution,
which include laser phase fluctuation and magnetic field fluctuation,
are incorporated into the equation as exponential decay of off-diagonal
density matrix elements for the internal degrees of freedom.
For the axial motional state, the initial distribution is 
assumed to be a thermal distribution based on the experimental results
of sideband cooling ($\overline{n_z}\sim0.02$).
Heating during the gate operation is neglected, which is reasonable
since our measured heating rate is $\sim$0.005 quanta/ms
for the axial motion and the typical gate sequences are shorter than 1 ms.

The parameters for the exponential dephasing are 
extracted from experimental results by manually fitting
simulation results for simple one-pulse sequences to
the experimental Rabi oscillation results.
Dotted curves in Fig.\ \ref{fig:rabi} represent such manually fitted
simulation results.
Representing exponential decay of off-diagonal density matrix elements
using a proportionality factor $\exp[-(\gamma/2){}t]$,
the values of $\gamma$ are extracted to be 
$2\pi\times400$ Hz for 
$\left|g\right\rangle$--$\left|\uparrow\right\rangle$
and 
$2\pi\times300$ Hz for 
$\left|\uparrow\right\rangle$--$\left|\downarrow\right\rangle$.

Based on the above-mentioned assumptions,
the CZ gate experiment is simulated with 4 pulses assumed, and the
result is shown in the Fig.\ \ref{fig:czgate}(b) as curves.
It well reproduces the reduction of fringe contrasts
and also the negative offset in the case of $\left|1\right\rangle$
preparation without any fitting parameters.
It is presumable that the negative offset is caused by the infidelity in the
BSB excitation on 
$\left|g\right\rangle$--$\left|\uparrow\right\rangle$
for the first and third pulse.
The infidelity result in population left in $\left|g\right\rangle$, 
which is considered as a leakage out of the computational
subspace spanned by 
$\left|\uparrow\right\rangle$, $\left|\downarrow\right\rangle$
leading to a signal decrease.

%\section{Discussions}

%\subsection{gate fidelity}
\label{sec:gate_fidelity}
The conditional phase shift confirmed above
is an essential result characterizing the CZ gate scheme,
while the whole functionality of the gate operation may be examined
more thoroughly 
through constructing truth tables,
performing quantum state/process tomography\cite{Riebe2006},
and estimating fidelity in entangled state generation using the gate.
Here we numerically simulate Bell state generation using the CZ gate
and estimate the fidelity.

A Bell state 
$\left|\Psi_B\right\rangle=(1/\sqrt{2})(\left|0\right\rangle\left|\uparrow\right\rangle+\left|1\right\rangle\left|\downarrow\right\rangle)$
can be produced from an initial state $\left|g\right\rangle$ by 
first applying a $\pi/2$ carrier pulse and a $\pi$ BSB pulse on
$\left|g\right\rangle$--$\left|\uparrow\right\rangle$
to prepare 
$(1/\sqrt{2})(\left|0\right\rangle+\left|1\right\rangle)\left|\uparrow\right\rangle$
and then performing a controlled NOT operation that flips the internal
qubit conditioned on the motional state.
Using exactly the same parameters as used for the simulation 
in Fig.\ \ref{fig:czgate}(b),
the time dependence of the density matrix 
in the process of the generation of the Bell state
$\left|\Psi_B\right\rangle$
is simulated.
The fidelity of the final state
$F\equiv\left\langle\Psi_B\right| \rho \left|\Psi_B\right\rangle$
is obtained to be 0.74,
which well exceed the value 0.5 expected for product states\cite{Sackett2000}.

Loss of fidelity in the Bell state generation process 
can be also estimated by simulation. 
Infidelity in excitation of the stimulated Raman transition
$\left|\uparrow\right\rangle$--$\left|\downarrow\right\rangle$,
which include phase noise between the lasers and magnetic field
fluctuation, contributes 12-14\%.
Infidelity in excitation of 
$\left|g\right\rangle$--$\left|\uparrow\right\rangle$,
which include 729-nm laser frequency noise and magnetic field
fluctuation, contributes 5-7\%.
Axial quantum number distribution contributes 6-7\%.
Intensity fluctuation is estimated to contribute by as low as 0.1\%,
and the effect of spontaneous emission is $\sim$0.1\%.

In conclusion, 
a quantum gate is demonstrated
with an atomic qubit consisting of electronic excited
states with a separation on the order of a terahertz.
A conditional dynamics is clearly observed as a fringe reversal
in Ramsey interferometry.
Fidelity for Bell-state generation is estimated
to be 0.74, and decoherence factors are analysed.
The excitation scheme using stimulated Raman transitions
with phase-locked lasers offers much flexibility, and 
is eventually used for atomic transitions which are not explored before
as qubit transitions as well as for molecular transitions. 

We thank I. Chuang and his group for their cooperation in the
development of the pulse programming system.
%\section{References}

%\bibliography{toyoda_et_al}

\end{document}